\documentstyle[12pt]{article}

\centerline{\bf  {\Large (2,0)-Super-Yang-Mills}}
\centerline {\bf{\Large Coupled to Non-Linear $\sigma$-Models}}
\vspace{0.7cm}
\centerline{M. S. G\'oes-Negr\~ao${^{(*)\dag}}$\footnote{negrao@cbpf.br
and mauro.negrao@inf.ucp.br},
 M. R.
Negr\~ao${^{(*)\dag}}$\footnote{guida@cbpf.br and
margarida.negrao@inf.ucp.br}} \centerline{ and A. B.
Penna-Firme${^{(+)\dag}}$\footnote{andrepf@cbpf.br}}
 \vspace{.5cm}

\centerline{$^{\dag}$Centro Brasileiro de Pesquisas F\'{\i}sicas- CBPF/CNPq }
\centerline{$^{*}$Universidade Cat\'olica de Petr\'opolis-GFT}
\centerline{$^{(+)}$Faculdade de Educa\c{c}\~ao da Universidade Federal
do Rio de Janeiro}

{
\begin{document}

\hspace\parindent

\begin{abstract}
\vspace{0.3cm}

{\footnotesize{\bf Considering a class of (2,0)-super-Yang-Mills
multiplets that accommodate a pair of independent gauge potentials
in connection with a single symmetry group, we present here their
coupling to ordinary matter and to non-linear $\sigma$-models in
(2,0)-superspace. The dynamics and the couplings of the gauge
potentials are discussed and the interesting feature that comes out
is a sort of ``chirality'' for one of the gauge potentials once
light-cone coordinates are chosen.}}
\end{abstract}

PACS numbers: 11.10.Lm, 11.15.-q and 11.30.Pb

\newpage

The raise of interest on the investigation of geometrical aspects
and quantum behaviour of two-dimensional systems, such as Yang-Mills
theories and non-linear $\sigma$-models, especially if endowed with
supersymmetry, has been broadly renewed in connection with the
analysis of superstring background configurations
\cite{{hull},{candelas}} and the study of conformal field theories
and integrable models.

As for supersymmetries defined in two space-time dimensions, they
may be generated by $p$ left-handed and $q$ right-handed independent
Majorana charges: these are the so-called $(p,q)$-supersymmetries
\cite{{hull},{saka}} and are of fundamental importance in the
formulation of the heterotic superstrings \cite{gross}.

Motivated by the understanding of a number of features related to
the dynamics of world-sheet gauge fields \cite{porrati} and the
possibility of finding new examples of conformal field theories,
one has considered the superspace formulation of a $(2,0)$-Yang-Mills model
\cite{{chair},{almeida}} enlarged by the introduction of an extra
gauge potential that transforms under the same {\it simple} gauge group
along with the ordinary Yang-Mills field of the theory.

In the works of refs.\cite{{chair},{almeida}}, one has discussed the
r\^ole of the further gauge potential on the basis of the
constraints upon field-strength superfields in the algebra of
gauge-covariant derivatives in $(2,0)$-superspace. The minimal
coupling of this somewhat less-constrained Yang-Mills model to matter
superfields has been contemplated, and it has been ascertained that
the additional gauge potential corresponds to non-interacting
degrees of freedom in the Abelian case. For non-Abelian symmetries,
the extra Yang-Mills field still decouples from matter, though it
presents self-interactions with the gauge sector \cite{chair}.

It is therefore our purpose in this letter to find out a possible
dynamical r\^ole for the additional gauge potential discussed in
refs.\cite{{chair},{almeida}}, by means of its coupling to matter
superfields that describe the coordinates of the K\"ahler manifold
adopted as the target space of a $(2,0)$ non-linear $\sigma$-model
\cite{dine}. To pursue such an investigation, we shall gauge the
isometry group of the $\sigma$-model under consideration, while
working in $(2,0)$-superspace; then, all we are left with is the
task of coupling the $(2,0)$-Yang-Mills extended supermultiplets of
ref.\cite{chair} to the superfields that define the $(2,0)$
$\sigma$-model whose gaugind is carried out.

The coordinates we choose to parametrise the $(2,0)$-superspace are
given by
\begin{eqnarray}
z^A
\equiv (x^{++}, x^{--}; \theta, \bar
\theta),
\label{coord}
\end{eqnarray}
where $x^{++}$, $x^{--}$ denote the usual light-cone variables,
whereas $\theta$, $\bar \theta$ stand for complex right-handed Weyl
spinors. The supersymmetry covariant derivatives are taken as:
\begin{eqnarray}
D_{+} \equiv \partial_{{\theta}} + i \bar \theta \partial_{++}
\end{eqnarray}
and
\begin{eqnarray}
\bar D_{+} \equiv \partial_{{\bar \theta}}
+ i \theta \partial_{++},
\label{susycode}
\end{eqnarray}
where $\partial_{++}$ (or $\partial_{--}$) represents the derivative
with respect to the space-time coordinate $x^{++}$ (or $x^{--}$).
They fulfill the algebra:
\begin{eqnarray}
D_{+}^{2} = {\bar D}_{+}^{2} = 0 \hspace{2.0cm}
\{ D_{+},{\bar D}_{+} \} = 2i
\partial_{++}.
\label{dquad}
\end{eqnarray}
With this definition for $D$ and ${\bar D}$, one can check that:
\begin{eqnarray}
e^{i{\theta}{\bar\theta}{\partial_{+}}}D_{+}e^{-i{\theta}{\bar\theta}{\partial_{+}}}
= {\partial_{\theta}},
\end{eqnarray}
\begin{eqnarray}
e^{-i{\theta}{\bar\theta}{\partial_{+}}}{\bar
D}_{+}e^{i{\theta}{\bar\theta}{\partial_{+}}} =
{\partial_{\bar\theta}}.
\end{eqnarray}

The fundamental matter superfields we shall deal with are the
``chiral'' scalar and left-handed spinor superfields, whose respective
component-field expressions are given by:
\begin{eqnarray}
{\Phi}(x;\theta,\bar \theta)&=&
e^{i\theta{\bar\theta}{\partial}_{++}}(\phi +\theta
\lambda),\nonumber\\ {\Psi}(x;\theta,\bar \theta)&=&
e^{i\theta{\bar\theta}{\partial}_{++}}(\psi +\theta \sigma);
\label{c5}
\end{eqnarray}
$\phi$ and $\sigma$ are scalars, whereas $\lambda$ and $\psi$
stand respectively for right- and left-handed Weyl spinors.

This sort of chirality constraint yields the following
component-field expansions for ${\Phi}^{i}$ and $\Psi^{i}$:
\begin{eqnarray}
{\Phi}^{i}(x;\theta,{\bar \theta}) &=& {\phi}^{i}(x) + \theta
{\lambda}^{i}(x) + i \theta {\bar
\theta}{\partial}_{++}
{\phi}^{i}(x),\nonumber\\ {\Psi}^{i}(x;\theta,{\bar \theta})&=&
{\psi}^{i}(x) + \theta {\sigma}^{i}(x) + i \theta {\bar
\theta}{\partial}_{++}
{\psi}^{i}(x).
\label{compon}
\end{eqnarray}

The most general superspace action involving $\Phi$ and $\Psi$ with
interactions governed by dimensionless coupling parameters
$f_1$ and $f_2$, reads
\begin{eqnarray}
S&=&\int d^2 x d\theta d{\bar \theta}[i({\bar
\Phi}{\partial}_{--}\Phi - \Phi\partial_{--}{\bar \Phi}) +{\bar
\Psi}\Psi +\nonumber\\ &+& m(\Phi \Psi + {\bar \Phi}\Psi)
+\nonumber\\ &+&f_1 P(\Phi,\bar \Phi)({\bar
\Phi}{\partial}_{--}\Phi - \Phi\partial_{--}{\bar
\Phi})+\nonumber\\ &+&f_2 Q(\Phi,\bar \Phi){\bar \Psi}\Psi],
\label{c7}
\end{eqnarray}
where $m$ is a mass parameter, and $P$ and $Q$ denote real
polynomials in $\Phi$ and ${\bar \Phi}$.

We now assume that both $\Phi$ and $\Psi$ transform under an arbitrary
compact and simple gauge group, ${\cal G}$, according to
\begin{eqnarray}
{\Phi}'=R(\Lambda)\Phi, \hspace{1cm} {\Psi}'=S(\Lambda)\Psi,
\label{c8}
\end{eqnarray}
where $R$ and $S$ are matrices that respectively represent a gauge
group element in the representations under which $\Phi$ and $\Psi$
transform. Taking into account the constraint on $\Phi$ and $\Psi$,
and bearing in mind the exponential representation of $R$ and $S$,
we find that the gauge parameter superfields, $\Lambda$, satisfy the
same sort of constraint. It can therefore be expanded as follows:
\begin{eqnarray}
{\Lambda}(x;\theta,\bar \theta)&=&
e^{i\theta{\bar\theta}{\partial}_{++}}(\alpha +\theta \beta),
\label{c9}
\end{eqnarray}
where $\alpha$ is a scalar and $\beta$ is a right-handed spinor.

The kinetic part of the action (\ref{c7}) can be made invariant
under the local transformations (\ref{c8}) by minimally coupling
gauge potential superfields, $\Gamma_{--}(x;\theta,{\bar \theta})$
and $V(x;\theta,{\bar \theta})$, according to the minimal coupling
prescriptions:
\begin{eqnarray}
S_{inv}=\int d^2 x d\theta d{\bar \theta} \{i[{\bar \Phi}e^{hV}
(\nabla_{--} \Phi)- ({\bar\nabla}_{--}
\bar\Phi)e^{hV}\Phi]+{\bar\Psi}e^{hV}\Psi\},
\label{c10}
\end{eqnarray}
where the gauge-covariant derivatives are define in the sequel.

The Yang-Mills supermultiplets are introduced by means of the
gauge-covariant derivatives which, according to the discussion of ref.
\cite{chair}, are defined as below:
\begin{eqnarray}
{\nabla}_{+} &\equiv & D_{+} + {\Gamma}_{+},\hspace{4.0cm} {\bar
\nabla}_{+} \equiv {\bar D}_{+},
\\
{\nabla}_{++} &\equiv & {\partial}_{++} + {\Gamma}_{++}
\hspace{1.5cm} and \hspace{1.5cm} {\nabla}_{--} \equiv
{\partial}_{--} -ig{\Gamma}_{--},
\label{dercov}
\end{eqnarray}
with the gauge superconnections ${\Gamma}_{+}$, ${\Gamma}_{++}$ and
${\Gamma}_{--}$ being all Lie-algebra-valued. The gauge couplings, $g$
and $h$, can in principle be taken different; nevertheless, this would \underline{not}
mean that we are gauging two independent symmetries. This is a single simple gauge group,
\cal{G}, with just one gauge-superfield parameter, ${\Lambda}$. It is the particular form of the
$(2,0)$-minimal coupling (realized by the exponentiation of ${V}$ and the connection present
in ${\nabla}_{--}$) that opens up the freedom to associate different
coupling parameters associated to $V$ and ${\Gamma}_{--}$. ${\Gamma}_{+}$ and
${\Gamma}_{++}$ can be both expressed in terms of the real scalar
superfield, $V(x;\theta,{\bar \theta})$, according to:
\begin{eqnarray}
{\Gamma}_{+} = e^{-gV}(D_{+} e^{gV})
\label{gamamais}
\end{eqnarray}
and
\begin{eqnarray}
{\Gamma}_{++}
= - \frac{i}{2} {\bar D}_{+} [e^{-gV}(D_{+}
e^{gV})].
\label{gama++}
\end{eqnarray}
Therefore, the gauging of the $\sigma$-model isometry group shall be
achieved by minimally coupling the action of the
$(2,0)$-supersymmetric $\sigma$-model to the gauge superfields $V$
and ${\Gamma}_{--}$, as we shall see in what follows.

To stablish contact with a component-field formulation and to
actually identify the presence of an aditional gauge potential, we
write down the $\theta$-expansions for $V$ and ${\Gamma}_{--}$:
\begin{eqnarray}
V(x;\theta,{\bar \theta}) = C + \theta
\xi - {\bar \theta}{\bar \xi} + \theta{\bar
\theta}v_{++}
\label{ve}
\end{eqnarray}
and
\begin{eqnarray}
{\Gamma}_{--}(x;\theta,{\bar
\theta}) &=& \frac{1}{2}(A_{--} + iB_{--}) + i\theta (\rho + i\eta)
\nonumber\\ &+& i{\bar \theta}
 (\chi + i\omega) + \frac{1}{2}\theta{\bar
\theta}(M+iN).
 \label{gama--}
 \end{eqnarray}

$A_{--}$, $B_{--}$ and $v_{++}$ are the light-cone components of the
gauge potential fields; $\rho, \eta, \chi$ and $\omega$ are
left-handed Majorana spinors; $M, N$ and $C$ are real scalars and
$\xi$ is a complex right-handed spinor.

It can be shown that the gauge transformations of the $\theta$-component fields above read
as follows:
\begin{eqnarray}
\delta C &=&\frac{2}{h} {\Im m}\alpha, \nonumber\\
\delta \xi &=& -\frac{i}{h}\beta, \nonumber\\
\delta v_{++}&=&\frac{2}{h} {\partial}_{++} \Re e\alpha, \nonumber\\
\delta A_{--}&=&\frac{2}{g} {\partial}_{--} \Re e\alpha, \nonumber\\
\delta B_{--}&=&\frac{2}{g} {\partial}_{--} {\Im m}\alpha, \nonumber\\
\delta \eta &=& -\frac{1}{g} {\partial}_{--} \Re e\beta, \nonumber\\
\delta \rho &=& \frac{1}{g} {\partial}_{--} {\Im m}\beta,\nonumber\\
\delta M &=& -\frac{2}{g} {\partial}_{++}{\partial}_{--} {\Im m}\alpha,\nonumber\\
\delta N &=& \frac{2}{g} {\partial}_{++}{\partial}_{--} \Re e\alpha, \nonumber\\
\delta \chi &=& 0, \nonumber\\
\delta \omega &=& 0,
\label{t}
\end{eqnarray}
and they suggest that we may take $h=g$, so that the
$v_{++}$-component
should be identified as the light-cone partner of $A_{--}$,
\begin{eqnarray}
v_{++}
\equiv A_{++};
\label{ve++}
\end{eqnarray}
this procedure yields two component-field gauge potentials: $A^{\mu}
\equiv (A^{0}, A^{1})$ and $B_{--}(x)$.

At this point, we should
 set a non-trivial remark: the $\theta {\bar \theta}$-component
 of $\Gamma_{--}$ should be identified as below:
\begin{eqnarray}
M+iN = i\partial_{++}(A_{--}+iB_{--}),
\label{21}
\end{eqnarray}
so as to ensure that $F_{\mu \nu}\equiv {\partial}_{\mu}
A_{\nu}-{\partial}_{\nu} A_{\mu}$ appear as a component accomodated
in the field-strength superfield defined by:
\begin{eqnarray}
[{\nabla}_{+},{\nabla}_{--}]\equiv X =-ig D_{+}{\Gamma}_{--}
-{\partial}_{--} \Gamma_{+}.
\label{fs}
\end{eqnarray}
$M$ and $N$ do not correspond therefore to independent degree of
freedom, and they are from now on suppressed from our
considerations.
The identification of eq.(\ref{21} does \underline{not} break supersymmetry,
for $\chi$ and $\omega$ are
non-dynamical degrees of freedom and drop out from the
field-strength superfield $X$. In practice, once this identification
has been adopted, $\Gamma_{--}$ carries two bosonic and
two fermionic degrees of freedom.

Using the field-strength defined in (\ref{fs}), we can build up the
gauge invariant kinetic Lagrangian:
\begin{eqnarray}
S_{kin} = -\frac{1}{8g^2}\int d^2 x d\theta d{\bar \theta} {\bar
X}X.
\end{eqnarray}

This action yields the component-field Lagrangian below:
\begin{eqnarray}
{\cal L}_{kin} = {\cal L}_{kin}(\rho,\eta,\xi) + {\cal
L}_{kin}(A)+{\cal L}_{kin}(B_{--},C),
\label{lk}
\end{eqnarray}
where
\begin{eqnarray}
{\cal L}_{kin}(\rho,\eta,\xi)= \frac{i}{8}({\bar \rho}-i{\bar
\eta} -{\partial}_{--} {\bar \xi})
\stackrel{\mathrm{\leftrightarrow}}{{\partial}}_{++}(\rho +i\eta -
{\partial_{--}}\xi),
\label{24}
\end{eqnarray}
with $A\stackrel{\mathrm{\leftrightarrow}}{{\partial}}B=(\partial
A)B-A(\partial B)$,
\begin{eqnarray}
{\cal L}_{kin}(A)=\frac{1}{2}A^{\nu}(\Box
\eta_{\mu\nu}-\partial_{\mu}\partial_{\nu})A^{\mu}=\frac{1}{2}A^{\nu}
R_{\mu\nu}A^{\mu},
\label{R}
\end{eqnarray}
and
\begin{eqnarray}
{\cal
L}_{kin}(B_{--},C)=\frac{1}{8}(\partial_{++}B_{--}-\partial_{++}\partial_{--}C)^{2}=\frac{1}{8}(B_{--}\hspace{0.2cm}C)K(B_{--}\hspace{0.2cm}C)^{t},
\label{K}
\end{eqnarray}
where the superscript $t$ stands for transposition.
Notice that, as already mentioned above, $\chi$ and $\omega$ are
\underline{not} present in the kinetic Lagrangian (\ref{24}).

Next, we can see that $R_{\mu\nu}$ and $K$ are singular matrices, so
it is necessary to write down a gauge fixing Lagrangian, which is
given by:
\begin{eqnarray}
S_{gf}&=& \frac{1}{8\alpha}\int d^2 x d\theta d\bar\theta {\bar G} G \nonumber\\
&=&-\frac{1}{2\alpha}(\partial_{\mu}A^{\mu})^{2}-\frac{i}{4\alpha}
({\bar \rho}-i{\bar \eta} -{\partial}_{--} {\bar
\xi}){\partial}_{++} (\rho +i\eta -
{\partial_{--}}\xi)+\nonumber\\
&-&\frac{1}{8\alpha}({\partial}_{++}B_{--} +
{\partial}_{--}{\partial}_{++}C)^{2},
\label{GF}
\end{eqnarray}
where $G=D_{+}\partial_{--}V-iD_{+}\Gamma_{--}$. Using the gauge-fixing,
eq.(\ref{GF}), along  with equations (\ref{R}) and
(\ref{K}), we are ready to write down the propagators for $A$, $B_{--}$ and
$C$:
\begin{eqnarray}
\langle AA \rangle &=& -\frac{2i}{\Box} (\theta^{\mu\nu} + \alpha \omega^{\mu\nu}),\nonumber\\
\langle BB \rangle &=& -\frac{8i}{{\Box}^2} (\alpha -1)\partial_{--}^2,\nonumber\\
\langle BC \rangle &=& -\langle CB\rangle =\frac{8i}{{\Box}^2} (\alpha + 1)\partial_{--},\nonumber\\
\langle CC \rangle &=& \frac{8i}{{\Box}^2} (\alpha - 1).
\label{prop}
\end{eqnarray}
The $C$-field exhibits a compensating character, as eqs.(\ref{t})
indicate. On the other hand, the field redefinition ${\tilde{B}}_{--}
\equiv B_{--}-\frac{h}{g}{\partial_{--}}C$ allows us to suppress $C$
from the action. In a way or another, $C$ is shown to be
non-physical. Here, we take the viewpoint to keep $C$ as
compensating; its elimination is accomplished by choosing the
$(2,0)$-Wess-Zumino-gauge, rather than upon the field reshuffling $B_{--} \mapsto
{\tilde{B}}_{--}$.
Expressing the action of equation (\ref{c10}) in terms of
component fields, and adopting the $(2,0)$-version of the Wess-Zumino gauge, the
matter-gauge sector Lagrangian reads:
reads as bellow:
\begin{eqnarray}
{\cal L}_{matter-gauge}&=& 2 \phi \Box {\phi}^{\ast} -ig A_{--}[{\phi}^{\ast}
{\partial}_{++}{\phi}-c.c] - ig A_{++}[{\phi}^{\ast}
{\partial}_{--}{\phi}-c.c]+\nonumber\\ &+& g \phi
{\phi}^{\ast}{\partial}_{++}B_{--} - g^{2} A_{++} A_{--}\phi{\phi}^{\ast}
+ 2i{\bar\lambda} {\partial}_{--}\lambda + g A_{--}{\bar
\lambda}\lambda +\nonumber\\
&+& -ig {\phi}^{\ast}[(\chi +{\bar \rho} +i\omega - i{\bar
\eta})\lambda
-c.c] -2i{\bar\psi}\partial_{++}\psi - gA_{+}{\bar\psi}\psi
+\nonumber\\
&+& {\bar{\sigma}}\sigma.
\end{eqnarray}
One immediately checks that the extra gauge field, $B_{--}$, does
\underline{not}
decouple from the matter sector. Our point of view of leaving the
superconnection $\Gamma_{--}$ as a complex superfield naturally
introduced this extra gauge potential in addition to the
usual gauge field $A_{\mu}$. $B_{--}$ behaves as a second gauge field.
The fact that it yelds a massless pole of order two in the spectrum may harm the
unitarity. However, the mixing with the $C$-component of $V$, which
is a compensating field, indicates that we should couple them to
external currents and analyse the imaginary part of the
current-current amplitude at the pole. In so doing, this imaginary
part turns out to be positive-definite, and so no ghosts are
present. This ensures us to state that $B_{--}$ behaves as a
physical gauge field: it has dynamics and couples to matter. Its
only peculiarity regards the presence of a single component in the
light-cone coordinates. The $B$-field plays rather the r\^ole of a
``chiral gauge potential''. Despite the presence of the pair of
gauge fields, a gauge-invariant mass term cannot be introduced, since
$B$ does not carry the $B_{++}$-component, contrary to what
happens with $A^{\mu}$. Let us now turn to the coupling of the two
gauge potentials, $A_{\mu}$ and $B_{--}$, to a non-linear
$\sigma$-model.

It is our main purpose henceforth to carry out the coupling of a
$(2,0)$ $\sigma$-model to the relaxed gauge superfields of the ref.
\cite{chair}, and show that the extra vector-degrees of freedom do not
decouple from the matter fields (that is the target space
coordinates). The extra gauge potential obtained upon relaxing
constraints can therefore acquire a dynamical significance by means
of the coupling between the $\sigma$-model and the Yang-Mills fields
of ref.\cite{chair}. Moreover, this system might provide another
example of a gauge-invariant conformal field theory.

The $(2,0)$-supersymmetric $\sigma$-model action written in
$(2,0)$-superspace reads \cite{dine}:
\begin{eqnarray}
S = - \frac{i}{2}
\int d^{2}x d\theta d {\bar \theta} \biggl [ K_{i}(\Phi, {\bar \Phi})
\partial_{--} {\Phi}^{i} - c.c. \biggl ]
,
\label{acao}
\end{eqnarray}
where the target space vector $K_{i}(\Phi, {\bar \Phi})$ can be
expressed as the gradient of a real scalar (K\"ahler) potential,
$K(\Phi,{\bar \Phi})$, whenever the Wess-Zumino term is absent ({\it
i.e.}, torsion-free case) \cite{hull}:
\begin{eqnarray}
K_{i}(\Phi, {\bar \Phi}) = {\partial}_{i} K(\Phi,{\bar \Phi})
\equiv
\frac{\partial}{\partial {\Phi}^{i}} K(\Phi, {\bar
\Phi}).
\end{eqnarray}

We shall draw our attention to K\"ahlerian target manifolds of the
coset type, $G/H$. The generators of the isometry group, $G$, are
denoted by $Q_{\alpha}(\alpha = 1,...,dim G)$, whereas the isotropy
group, $H$, has its generators denoted by $Q_{\bar \alpha}(\bar
\alpha = 1,...,dim H)$. The transformations of the isotropy group are
linearly realised on the superfields $\Phi$ and $\bar \Phi$, and act
as matrix multiplication, just like for flat target manifolds. The
isometry transformations instead are non-linear, and their
infinitesimal action on the points of $G/H$ can be written as:
\begin{eqnarray}
\delta {\Phi}^{i}
=
{\lambda}^{\alpha}k_{\alpha}^{i}(\Phi)
\label{infact}
\end{eqnarray}
and
\begin{eqnarray}
\delta
{\bar \Phi}_{i} = {\lambda}^{\alpha}{\bar k}_{\alpha i}(\bar \Phi),
\label{infactbar}
\end{eqnarray}
where $k_{\alpha i}$ and ${\bar k}_{\alpha i}$ are respectively
holomorphic and anti-holomorphic Killing vectors of the target
manifold. The finite versions of the isometry transformations above
read:
\begin{eqnarray}
{\Phi}'^{i} = exp({\bf L}_{\lambda . k}){\Phi}^{i}
\end{eqnarray}
and
\begin{eqnarray}
{\bar \Phi}'^{i} = exp({\bf L}_{\lambda . {\bar k}}){\bar
\Phi}^{i}
\end{eqnarray}
with
\begin{eqnarray}
{\bf L}_{\lambda . k} {\Phi}^{i} \equiv \biggl [
{\lambda}^{\alpha}k_{\alpha}^{i}
\frac{\partial}{\partial {\Phi}^{j}}, {\Phi}^{i} \biggl ] =
\delta{\Phi}^{i}.
\end{eqnarray}

Though the K\"ahler scalar potential can always be taken
$H$-invariant, isometry transformations induce on $K$ a variation
given by:
\begin{eqnarray}
\delta K = {\lambda}^{\alpha} [
({\partial}_{i} K) k_{\alpha i} + ({\bar \partial}^{i} K){\bar
k}_{\alpha i} ] = {\lambda}^{\alpha} [ {\eta}_{\alpha}(\Phi) + {\bar
\eta}_{\alpha} (\bar \Phi) ],
\label{transK}
\end{eqnarray}
where the holomorphic and anti-holomorphic functions
${\eta}_{\alpha}$ and ${\bar \eta}_{\alpha}$ can be determined up to
a purely imaginary quantity as below:
\begin{eqnarray}
({\partial}_{i} K)k_{\alpha}^{i} \equiv {\eta}_{\alpha} + i
M_{\alpha}(\Phi,{\bar
\Phi})
\end{eqnarray}
and
\begin{eqnarray}
({\bar \partial}^{i} K) {\bar k}_{\alpha i} \equiv {\bar \eta}_{\alpha} - i
M_{\alpha}(\Phi, {\bar
\Phi}).
\end{eqnarray}

The real functions $M_{\alpha}$, along with the holomorphic and
anti-holomorphic functions ${\eta}_{\alpha}$ and ${\bar
\eta}_{\alpha}$, play a crucial r\^ole in discussing the gauging of the
isometry group of the target manifold \cite{{bagger},{kar}}.
Therefore, by virtue of the transformation (\ref{transK}) and the
constraints imposed on $\Phi$ and $\bar \Phi$, it can be readily
checked that the superspace action (\ref{acao}) is {\it invariant}
under global isometry transformations.

Proceeding further with the study of the isometries, a relevant
issue in the framework of $(2,0)$-supersymmetric $\sigma$-models is
the gauging of the isometry group $G$ of the K\"ahlerian target
manifold. This in turn means that one should contemplate the minimal
coupling of the $(2,0)$-$\sigma$-model to the Yang-Mills
supermultiplets of $(2,0)$-supersymmetry \cite{brooks}. An eventual
motivation for pursuing such an analysis is related to the
$2$-dimensional conformal field theories. It is known that
$2$-dimensional $\sigma$-models define conformal field theories
provided that suitable constraints are imposed upon the target space
geometry \cite{{hull},{candelas}}. Now, the coupling of these models
to the Yang-Mills sector might hopefully yield new conformal field
theories of interest.

The study of $(2,0)$-supersymmetric Yang-Mills theories has been
carried out in ref.\cite{brooks} and the gauging of $\sigma$-model
isometries in $(2,0)$-superspace has been considered in
ref.\cite{hel}. On the other hand, our alternative less-constrained
version of $(2,0)$-gauge multiplets indicates that the
elimination of some constraints on the gauge superconnections and on
field-strength superfields leads to the appearence of an extra gauge
potential that shares a common gauge field in partnership with the usual Yang-Mills field.
We wish henceforth to analyse the coupling of our ``loose´´ $(2,0)$-multiplets to non-linar
$\sigma$-model.

To write down the local version of the isometry transformations
(\ref{infact}) and (\ref{infactbar}), we have to replace the global
parameter ${\lambda}^{\alpha}$ by a pair of chiral and antichiral
superfields, $\Lambda^{\alpha}(x;\theta,{\bar \theta})$ and ${\bar
\Lambda}^{\alpha}(x;\theta,{\bar \theta})$, by virtue of the
constraints satisfied by $\Phi$ and ${\bar \Phi}$. This can be
realised according to:
\begin{eqnarray}
{\Phi}^{'i} = exp({\bf L}_{\Lambda . k}) {\Phi}^{i}
\label{filocal}
\end{eqnarray}
and
\begin{eqnarray}
{\bar
\Phi}^{'i} = exp({\bf L}_{\bar \Lambda . \bar k}) {\bar
\Phi}^{i}.
\label{barfilocal}
\end{eqnarray}

In order to get closer to the case of global transformations, and to
express all gauge variations exclusively in terms of the superfield
parameters ${\Lambda}^{\alpha}$, we propose a field redefinition
that consist in replacing ${\bar \Phi}$ by a new superfield,
${\tilde \Phi}$, as it follows:
\begin{eqnarray}
{\tilde
\Phi}_{i} \equiv exp(i{\bf L}_{V.\bar k}){\bar
\Phi}_{i}.
\end{eqnarray}
From the expression for the gauge transformation of the prepotential
$V$, it can be shown that:
\begin{eqnarray}
exp(i{\bf L}_{V'.{\bar k}}) = exp({\bf L}_{{\Lambda}.{\bar
k}})exp(i{\bf L}_{V.{\bar k}})exp(-{\bf L}_{{\bar\Lambda}{\bar k}}),
\end{eqnarray}
and ${\tilde\Phi_{i}}$ consequently transform with the gauge
parameter ${\Lambda^{\alpha}}$:
\begin{eqnarray}
{{\tilde\Phi}_{i}}' = exp({\bf L}_{{\Lambda}{\bar k}}){\tilde
\Phi}_{i},
\end{eqnarray}
which infinitesimally reads:
\begin{eqnarray}
{\delta{\tilde\Phi}}_{i} =
{\Lambda}^{\alpha}(x;{\theta}{\bar\theta}){\bar k}_{\alpha
i}(\bar\Phi).
\end{eqnarray}

Now, an infinitesimal isometry transformation induces on the
modified K\"ahler potential, $K({\Phi},{\tilde\Phi})$, a variation
given by:
\begin{eqnarray}
{\delta K}({\Phi},{\tilde \Phi}) =
{\Lambda}^{\alpha}({\eta}_{\alpha} + {\tilde
\eta}_{\alpha}),
\end{eqnarray}
where
\begin{eqnarray}
{\tilde\eta}_{\alpha} = ({\tilde\partial}^{i}K){\bar
k}_{{\alpha}i}(\tilde \Phi) + iM_{\alpha}({\Phi},{\tilde\Phi}),
\end{eqnarray}
with ${\tilde \partial}$ denoting a partial derivative with respect
to ${\tilde \Phi}$. The isometry variation ${\delta K}$ computed
above reads just like a K\"ahler transformation and is a direct
consequence of the existence of the real scalar
$M_{\alpha}({{\Phi},{\tilde \Phi}})$, as discussed in refs
\cite{{bagger},{kar}}.

The form of the isometry variation of $K({\Phi},{\tilde\Phi})$
suggests the introduction of a pair of {\it chiral} and {\it
antichiral} superfields, ${\xi(\Phi)}$ and ${\bar\xi}(\bar\Phi)$,
whose respective gauge transformations are fixed in such a way that
they compensate the change of K under isometries. This can be
achieved by means of the Lagrangian defined as:
\begin{eqnarray}
{\cal L}_{\xi} &=& {\partial}_{i}[K(\Phi,\tilde\Phi) - {\xi}(\Phi)
- {\tilde\xi}(\tilde\Phi)]{\nabla}_{--}{\Phi}^{i} + \nonumber\\ &-&
{\tilde\partial}_{i}[K(\Phi,\tilde\Phi) - {\xi}(\Phi) -
{\tilde\xi}(\tilde\Phi)]{\nabla}_{--}{\tilde\Phi}^{i},
\label{calL}
\end{eqnarray}
where the covariant derivatives ${\nabla}_{--}{\Phi}^{i}$ and
${\nabla}_{--}{\tilde \Phi}^{i}$ are defined in perfect analogy to
what is done in the case of the bosonic ${\sigma}$-model:
\begin{eqnarray}
{\nabla}_{--}{\Phi}_{i} \equiv {\partial}_{--}{\Phi}_{i} - g
{\Gamma}_{--}^{\alpha} {k}_{\alpha}^{i}(\Phi)
\end{eqnarray}
and
\begin{eqnarray}
{\nabla}_{--}{\tilde \Phi}_{i}
\equiv {\partial}_{--}{\tilde \Phi}_{i} - g {\Gamma}_{--}^{\alpha} {\bar
k}_{\alpha i} (\tilde \Phi).
\end{eqnarray}

Finally, all we have to do in order that the Lagrangian ${\cal
L}_{\xi}$ given above be invariant under local isometries is to fix
the gauge variations of the auxiliary scalar superfields $\xi$ and
${\tilde \xi}$. If the latter are so chosen that:
\begin{eqnarray}
({\partial}_{i}\xi)k_{\alpha}^{i}(\Phi) = {\eta}_{\alpha}(\Phi)
\end{eqnarray}
and
\begin{eqnarray}
({\tilde
\partial}^{i}{\tilde \xi}){\bar k}_{\alpha i} (\tilde \Phi) = {\tilde
\eta}_{\alpha} ({\tilde \Phi}),
\end{eqnarray}
then it can be readily verified that the K\"ahler-transformed
potential
\begin{eqnarray}
[K(\Phi,\bar \Phi) - \xi(\Phi) - {\tilde \xi}(\tilde \Phi)]
\end{eqnarray}
is isometry-invariant, and the Lagrangian ${\cal L}_{\xi}$ of eq.
(\ref{calL}) is indeed symmetric under the gauged isometry group.

The interesting point we would like to stress is that the extra
gauge degrees of freedom accommodated in the component-field
$B_{--}(x)$ of the superconnection ${\Gamma}_{--}$ behave as a
genuine gauge field that shares with $A^{\mu}$ the coupling to
matter and to $\sigma$-model \cite{chair}. This result can be
explicitly read off from the component-field Lagrangian projected
out of the superfield Lagrangian ${\cal L}_{\xi}$. We therefore
conclude that our less constrained $(2,0)$-gauge theory yields a
pair of gauge potentials that naturally transform under the action
of a single compact and simple gauge group.

The authors would like to thank J. A. Hellay\"el-Neto, M. A. De
Andrade and A. L. M. A. Nogueira for enlightening discussions and to
CNPq and Capes are acknowledged for the financial support.  M. S.
G\'oes-Negr\~ao and M. R. Negr\~ao are also grateful
to DF-UEFS where part of
this work was done.

\end{document}